# Bagging multiple comparisons from microarray data

Dimitris N. Politis[a]

[a]Department of Mathematics, University of California at San Diego, La Jolla, CA 92093-0112.

**ABSTRACT**
**Motivation:** Multiple comparisons arising from gene expression data are typically low power experiments. As a result, lists of genes declared non-null by different copies of the same experiment can be totally disjoint.
**Results:** Motivated by the above observation, the methodologies of bagging and subagging are put forth for the purpose of increasing the discovery power in multiple comparisons from microarray data without significantly affecting their false discovery rate.
**Availability:** n/a
**Contact:** dpolitis@ucsd.edu

## 1 INTRODUCTION

The problem of simultaneous statistical inference is not new; see Miller (1981) for an early treatment. In the last decade, however, the statistical community has been faced with huge amounts of data and a subsequent need to address large-scale simultaneous hypothesis testing problems.

The prototypical such dataset involves gene expression data but different applications, such as functional Magnetic Resonance Imaging, flight spectroscopy, flow cytometry, etc., all give rise to similar problems from a statistician's perspective. The microarray set-up is described below in the context of the gene expression example with the understanding that the same ideas are applicable to a host of other two-sample, multiple comparison problems.

A typical experiment may entail data on $n_X$ normal subjects, and $n_Y$ patients. An array of $N$ measurements is obtained from each subject. Therefore, the data can be organized as a $N \times n_X$ data matrix $X$ (control group), and a $N \times n_Y$ data matrix $Y$ (patient group); the $(i,j)$ entry of $X$ is denoted $X_{ij}$, and that of $Y$ is denoted $Y_{ij}$. Column $i$ from $X$ has the data from the $i$th normal subject, and column $j$ from $Y$ has the data from the $j$th patient.

The $X$ data are assumed independent of the $Y$ data. A general model for this set-up is to assume that, for each $k$, $X_{k,1}, X_{k,2}, \ldots, X_{k,n_X} \sim$ i.i.d. $F_X^{(k)}$, and $Y_{k,1}, Y_{k,2}, \ldots, Y_{k,n_Y} \sim$ i.i.d. $F_Y^{(k)}$ where $F_X^{(k)}, F_Y^{(k)}$ are some distribution functions. For each $k = 1, \ldots, N$, the issue is to test $H_0 : F_X^{(k)} = F_Y^{(k)}$ vs. not; this is the set-up of multiple comparisons.

More often than not, the testing focuses on a potential difference in the means of the $X$ and $Y$ data. In that case, practitioners typically assume

$$X_{k,1}, X_{k,2}, \ldots, X_{k,n_X} \sim \text{i.i.d. } N(\mu_k, \sigma_k^2) \quad (1)$$

and $\quad Y_{k,1}, Y_{k,2}, \ldots, Y_{k,n_Y} \sim \text{i.i.d. } N(\nu_k, \sigma_k^2). \quad (2)$

The multiple comparisons now boil down to testing $H_0 : \mu_k = \nu_k$ vs. not, for $k = 1, \ldots, N$. From the $k$th row, the familiar $t$-statistic $t^{(k)} = (\bar{Y}_{k\cdot} - \bar{X}_{k\cdot})/(\hat{\sigma}\sqrt{n_Y^{-1} + n_X^{-1}})$ can be calculated where $\bar{Y}_{k\cdot} = n_Y^{-1}\sum_{j=1}^{n_Y} Y_{kj}$, $\bar{X}_{k\cdot} = n_X^{-1}\sum_{i=1}^{n_X} X_{ki}$, and $\hat{\sigma}^2 = (n_X + n_Y - 2)^{-1}\{\sum_{i=1}^{n_X}(X_{ki} - \bar{X}_{k\cdot})^2 + \sum_{j=1}^{n_Y}(Y_{kj} - \bar{Y}_{k\cdot})^2\}$ is the pooled variance.[1] A typical testing procedure then rejects $H_0$ from the $k$th row when $t^{(k)}$ is too large in absolute value.

Suppose that exactly $n_0$ rows (genes) conform to $H_0$, i.e., they are "null", and so $N - n_0$ rows (genes) do not, i.e., they are "non-null". Collect the indices of the trully non-null rows in a list denoted by TRUELIST; similarly, collect the row indices corresponding to the rejected $t$-statistics in the LIST of genes declared to be non-null. Then we can define the multiple comparisons achieved discovery power as

$$ADP = \frac{\#\{LIST \cap TRUELIST\}}{\#\{TRUELIST\}}$$

and the achieved false discovery rate as

$$AFDR = \frac{\#\{LIST \cap \overline{TRUELIST}\}}{\#\{LIST\}}$$

where $\#\{A\}$ denotes number of elements in set $A$, and $\bar{A}$ is the complement of $A$. The breakthrough method of Benjamini and Hochberg (1995) was designed to control the expected value of the AFDR; this expected value is usually called simply the false discovery rate (FDR).

## 2 APPROACH

Suppose that two different groups perform the same scientific experiment and come up with two different lists of genes declared non-null, say $LIST_1$ and $LIST_2$. Let $AFDR_1$ and $AFDR_2$ denote the false discovery rates in the two experiments; recall that (the expected values of) $AFDR_1$ and $AFDR_2$ are controlled, i.e., bounded, in a typical multiple comparisons experiment.

How can the two lists, $LIST_1$ and $LIST_2$, be combined for better inference? The natural answer is to 'heed' the evidence from both experiments and declare as non-null all elements in the $BIGLIST = LIST_1 \cup LIST_2$. Since the BIGLIST is bigger than either $LIST_1$ or $LIST_2$, the combined experiment will have more power; but what is the AFDR associated with the BIGLIST?

---
[1] The normality assumption is not crucial in practice, especially if the sample sizes $n_X$ and $n_Y$ are relatively large. The assumption of common variance on the $k$th row of $X$ and $Y$ is more important but can be addressed if required leading to a slightly different form of the $t$-statistic; in any case, the flavor of the testing problem remains unchanged.

To proceed with the analysis, let us make the simplifying assumption that genes declared non-null in both studies are very likely trully non-null, i.e., that $SMALLLIST \subset TRUELIST$ with high probability where we denote $SMALLLIST = LIST_1 \cap LIST_2$. Also let $FALSE_1$ denote the subset of $LIST_1$ that consists of false discoveries, i.e., genes falsely declared non-null; similarly for $FALSE_2$. Therefore, we have:

$$AFDR_1 = \frac{\#\{FALSE_1\}}{\#\{LIST_1\}}, \quad AFDR_2 = \frac{\#\{FALSE_2\}}{\#\{LIST_2\}}, \quad (3)$$

from which the numbers $\#\{FALSE_1\}$ and $\#\{FALSE_2\}$ can be calculated as functions of $AFDR_1$ and $AFDR_2$.

Consequently, the AFDR associated with BIGLIST is given by:

$$AFDR_{BIG} = \frac{\#\{FALSE_1\} + \#\{FALSE_2\}}{\#\{LIST_1\} + \#\{LIST_1\} - \#\{SMALLLIST\}}$$

$$= \frac{AFDR_1 \times \#\{LIST_1\} + AFDR_2 \times \#\{LIST_2\}}{\#\{LIST_1\} + \#\{LIST_1\} - \#\{SMALLLIST\}}. \quad (4)$$

Taking expectations in the above, we see that eq. (4) is satisfied with the expected false discovery rates (FDR) in place of the AFDRs, i.e., that:

$$FDR_{BIG} = \frac{FDR_1 \times \#\{LIST_1\} + FDR_2 \times \#\{LIST_2\}}{\#\{LIST_1\} + \#\{LIST_1\} - \#\{SMALLLIST\}}. \quad (5)$$

In experiments with low power it is not uncommon to have $LIST_1$ and $LIST_2$ be totally disjoint; see Efron (2006) for a discussion. Suppose we are in such a low-power set-up, and also suppose—for the sake of argument—that the two experiments have similar design, i.e., that $FDR_1 = FDR_2$. Then, the above equations show that $FDR_{BIG} = FDR_1 = FDR_2$. So, in this case, the combined experiment has more power with the same FDR, i.e., a win-win situation.

In general, however, $LIST_1$ and $LIST_2$ might not be disjoint, and the increase in power associated with BIGLIST will come at the price of an increase in FDR. However, it is the thesis of this paper that the increase in power may be well worth a small increase in FDR.

Before proceeding further, let us momentarily consider the generalization to the case of having $M$ different groups perform the same experiment and coming up with their respective non-null lists, say $LIST_1, LIST_2, \ldots, LIST_M$; let $AFDR_1, AFDR_2 \ldots, AFDR_M$ denote the respective AFDRs. Under the same simplifying assumption, namely that genes declared non-null in at least two studies are very likely truly non-null, a similar calculation as before yields:

$$FDR_{BIG} = \frac{\sum_{i=1}^{M} FDR_i \times \#\{LIST_i\}}{\#\{BIGLIST\}} \quad (6)$$

where again $FDR_{BIG}$ is the expected false discovery rate associated with $BIGLIST = \cup_{i=1}^{M} LIST_i$. Finally, note that the number of elements in $BIGLIST$ can be calculated as:

$$\#\{BIGLIST\} = \sum_i \#\{LIST_i\} - \sum_{i \neq j} \#\{LIST_i \cap LIST_j\}$$

$$+ \sum_{i \neq j \neq k \neq i} \#\{LIST_i \cap LIST_j \cap LIST_k\} + \ldots$$

$$\ldots + (-1)^{M-1} \times \#\{\cap_{i=1}^{M} LIST_i\}.$$

## 3 METHODS

### 3.1 Bootstrap and bagging

In Section 2, having multiple experiments (with their associated rejection LISTs) was discussed. In practice, however, the statistician is faced with a single dataset. Nonetheless, resampling and subsampling methods can be utilised in order to create additional (pseudo)samples.

Efron's (1979) bootstrap is one of the most prominent resampling methods. For i.i.d. data $Z_1, \ldots, Z_n$, the bootstrap amounts to sampling randomly with replacement from the set $\{Z_1, \ldots, Z_n\}$ to create the (pseudo)sample $Z_1^*, \ldots, Z_n^*$; see Efron and Tibshirani (1993) for a review. The bootstrap is closely related to Tukey's (1958) 'delete-1' jackknife which was generalized to a 'delete-$d$' jackknife by Shao and Wu (1989). For i.i.d. data $Z_1, \ldots, Z_n$, the delete-$d$ jackknife is equivalent to subsampling with sample size $b = n - d$, i.e., sampling randomly without replacement from the set $\{Z_1, \ldots, Z_n\}$ to create the (pseudo)sample $Z_1^*, \ldots, Z_b^*$; see Politis, Romano and Wolf (1999).

'Bagging', i.e., bootstrap aggregation, was put forth by Breiman (1996) in order to improve the accuracy of statistical predictors. The idea is to evaluate the predictor in question on a number of bootstrap (pseudo)datasets, and to combine the resulting predictors in an aggregate predictor. It has been shown that bagging indeed helps improve predictor accuracy in particular when the predictor is relatively unstable, i.e., when small changes in the data result in greatly perturbed predictions; see Bühlmann and Yu (2002). Bagging can alternatively be implemented in conjunction with subsampling in which case the term 'subagging' was suggested by Bühlmann and Yu (2002); see also Bühlmann (2003).

### 3.2 Balanced bagging and subagging for microarrays

As discussed in Section 2, it is possible to have two different low-power experiments produce disjoint or almost disjoint rejection lists; this is evidence of instability. Thus, bagging and/or subagging may be helpful for multiple comparisons as they have been shown to be helpful in prediction and classification.

We now elaborate on how to perform bagging and subagging in the multiple comparisons, microarray set-up of Section 1; the main idea is to re/sub-sample subjects, i.e., columns of the matrices $X$ and $Y$. Throughout this section it is assumed that the practitioner is using a fixed multiple hypothesis testing procedure, e.g., the procedure of Benjamini and Hochberg (1995) or Efron (2005), for *any* dataset that he/she may encounter.

The bagging and subagging algorithms described below are termed 'balanced'; the reason for this term will become more apparent in Section 3.4. Let $\underline{x}_1, \ldots, \underline{x}_{n_X}$ and $\underline{y}_1, \ldots, \underline{y}_{n_Y}$ denote the columns of $X$ and $Y$ respectively; $B$ is an integer denoting the number of (pseudo)samples generated.

- Balanced bagging. For $k = 1, \ldots, B$, construct the $k$th bootstrap (pseudo)sample $X^{(k)}$ and $Y^{(k)}$; the columns of $X^{(k)}$ and $Y^{(k)}$ respectively are given as $\underline{x}_{I_1}, \ldots, \underline{x}_{I_{n_X}}$ and $\underline{y}_{J_1}, \ldots, \underline{y}_{J_{n_Y}}$ where $I_1, \ldots, I_{n_X}$ are numbers drawn randomly with replacement from the index set $\{1,\ldots,n_X\}$ and $J_1,\ldots,J_{n_Y}$ are numbers drawn randomly with replacement from the index set $\{1,\ldots,n_Y\}$ and independently of $I_1,\ldots,I_{n_X}$. From this $k$th (pseudo)sample, the rejection list $LIST_k$ is created.

To define subagging, subsample sizes $b_X$ and $b_Y$ must be specified. Note that there is no reason here to have the subsample sizes be of smaller order of magnitude as compared to the original sample sizes; this is only required for estimation consistency which is not the objective here—see e.g. Politis et al. (1999). So, the subsample sizes for subagging could (and should) be taken relatively large; furthermore, it is intuitive that a choice satisfying $b_X/b_Y \simeq n_X/n_Y$ might be fruitful as being more representative of the original dataset. Thus, a good rule-of-thumb may be to let $b_X \simeq a\, n_X$ and $b_Y \simeq a\, n_Y$ where the constant $a$ is close to (but less than) one.

- Balanced subagging—random version. For $k = 1,\ldots,B$, construct the $k$th subagging (pseudo)sample $X^{(k)}$ and $Y^{(k)}$; the columns of $X^{(k)}$ and $Y^{(k)}$ respectively are given as $\underline{x}_{I_1},\ldots,\underline{x}_{I_{b_X}}$ and $\underline{y}_{J_1},\ldots,\underline{y}_{J_{b_Y}}$ where $I_1,\ldots,I_{b_X}$ are numbers drawn randomly without replacement from the index set $\{1,\ldots,n_X\}$ and $J_1,\ldots,J_{b_Y}$ are numbers drawn randomly without replacement from the index set $\{1,\ldots,n_Y\}$ and independently of $I_1,\ldots,I_{b_X}$. As before, from this $k$th (pseudo)sample, the rejection list $LIST_k$ is created.

- Balanced subagging—nonrandom version. Let $\mathbf{S}_X$ denote the set of all size $b_X$ subsets of the index set $\{1,\ldots,n_X\}$, and $\mathbf{S}_Y$ denote the set of all size $b_Y$ subsets of the index set $\{1,\ldots,n_Y\}$ where $b_X$ and $b_Y$ are as above. A subagging (pseudo)sample is given by $X^{(k_1)}$ and $Y^{(k_2)}$ where the columns of $X^{(k_1)}$ are the columns of $X$ with indices given by the $k_1$th element of set $\mathbf{S}_X$, and the columns of $Y^{(k_2)}$ are the columns of $Y$ with indices given by the $k_2$th element of set $\mathbf{S}_X$. Since the set $\mathbf{S}_X$ contains $\frac{n_X!}{b_X!(n_X-b_X)!}$ elements and the set $\mathbf{S}_Y$ contains $\frac{n_Y!}{b_Y!(n_Y-b_Y)!}$ elements, it is apparent that there are $B = \frac{n_X!}{b_X!(n_X-b_X)!} \cdot \frac{n_Y!}{b_Y!(n_Y-b_Y)!}$ possible (pseudo)samples.

Of course, $\frac{n_X!}{b_X!(n_X-b_X)!} \cdot \frac{n_Y!}{b_Y!(n_Y-b_Y)!}$ can be a prohibitively large number, so considering all possible (pseudo)samples seems out of the question. The aforementioned random subagging procedure side-steps this difficulty but so does the following scheme that has the additional benefit of nonrandom selection of 'maximum contrast' subsamples, i.e., subsamples that are 'most' different from one another in their composition.

It is easier to describe this idea in the 'delete-$d$' framework (with $d = n - b$) as opposed to 'choose-$b$'; of course, now the game is delete-$d$ columns from one of our data matrices.

- 'Maximum contrast' nonrandom subagging. Let $m_X, m_Y$ be two positive integers, and divide the index set $\{1,\ldots,n_X\}$ into the $m'_X$ subsets $S_X^{(1)},\ldots,S_X^{(m'_X)}$ where $S_X^{(1)} = \{1,\ldots,d_X\}, S_X^{(2)} = \{d_X+1,\ldots,2d_X\},\ldots$, etc. where $d_X = \lceil n_X/m_X \rceil$ and $m'_X = \lceil n_X/d_X \rceil$; here $\lceil a \rceil$ is the smallest integer that is bigger or equal to $a$. The last set, i.e., $S_X^{(m'_X)}$, may have size less than $d_X$ if $m_X$ does not divide $n_X$ but that poses no problem. Similarly, divide the index set $\{1,\ldots,n_Y\}$ into the $m'_Y$ subsets $S_Y^{(1)},\ldots,S_Y^{(m'_Y)}$ where $S_Y^{(1)} = \{1,\ldots,d_Y\}, S_Y^{(2)} = \{d_Y+1,\ldots,2d_Y\},\ldots$, etc. A subagging (pseudo)sample is now given by $X^{(k_1)}$ and $Y^{(k_2)}$ where the columns of $X^{(k_1)}$ are the columns of $X$ with indices given by the set $\{1,\ldots,n_X\} - S_X^{(k_1)}$, and the columns of $Y^{(k_2)}$ are the columns of $Y$ with indices given by the set $\{1,\ldots,n_Y\} - S_Y^{(k_2)}$. Since the possible values of $k_1$ are $\{1,\ldots,m'_X\}$, and those for $k_2$ are $\{1,\ldots,m'_Y\}$, it is apparent that there are $m'_X \cdot m'_Y$ possible such (pseudo)samples; thus rejection lists $LIST_1,\ldots,LIST_B$ can be created with $B = m'_X \cdot m'_Y$.

### 3.3 Combining the rejection lists

Let $LIST$ denote the rejection list of the original dataset $X$ and $Y$, and $LIST_1,\ldots,LIST_B$ the rejections lists corresponding to $B$ (pseudo)samples from one of the algorithms of Section 3.2.

As in Section 2, the simplest suggestion is to combine the lists by a union, i.e., to define the aggregate/combined list as:

$$LIST.AGG = LIST \cup LIST1 \cup LIST2 \cup \cdots \cup LISTB. \quad (7)$$

However, other alternatives exist; their description is facilitated by the notion of 'voting' where a list is said to 'vote' that the $i$th gene is non-null when the $i$th gene is an element of the list.

Let $V(i)$ denote the number of votes the $i$th gene received from the 'voting' lists $LIST, LIST_1,\ldots,LIST_B$. With this terminology, rejecting every gene in $LIST.AGG$ corresponds to the formula:

(i) declare the $i$th gene as non-null if $V(i) \geq 1$, i.e., it got at least one vote.

A more conservative approach might require to 'second' a vote, i.e., it would

(ii) declare the $i$th gene as non-null if $V(i) \geq 2$, i.e., it got at least two votes.

One might even raise the rejection threshold at a level higher than two although we will not consider that here. However, it is informative to see which genes received more votes than others in the sense that getting more votes corresponds to more evidence for being truly non-null. Thus, a plot of $V(i)$ vs. $i$ may be a helpful diagnostic tool.

As a further diagnostic, we may define $N(h)$ as the number of genes that received at least $h$ votes, i.e., $N(h)$ is the size of the non-null list obtained from a criterion of the type: reject gene $i$ if $V(i) \geq h$. A plot of $N(h)$ vs. $h$ is another way to quantify the 'strength of evidence' towards proclaiming each gene on $LIST.AGG$ as non-null.

Note that formula (ii) treats $LIST$ as 'equal' to $LIST_1,\ldots,LIST_B$, and carries the implicit risk that not all of the genes found in $LIST$ will be finally rejected. To remedy this, we may give the original $LIST$ more weight in the aggregation. The easiest way of doing this is giving the original $LIST$ a double vote, i.e., defining $V^*(i)$ to equal the number of votes the $i$th gene got from $LIST_1,\ldots,LIST_B$ plus a double vote from the original $LIST$ (if indeed $LIST$ gave it a vote),[2] and then

(ii*) declaring the $i$th gene as non-null if $V^*(i) \geq 2$.

As above, we can define $N^*(h)$ as the number of genes that received at least $h$ votes from formula (ii*) above, i.e., $N^*(h)$ is the size of the non-null list obtained from a criterion of the type: reject gene $i$ if $V^*(i) \geq h$. A plot of $N^*(h)$ vs. $h$ has an interpretation similar to that of plot of $N(h)$ vs. $h$.

---

[2] Note that we can also get $V^*(i)$ by computing $V(i)$ counting single votes from $LIST, LIST$, and $LIST_1,\ldots,LIST_B$, i.e., having $LIST$ double-up and—in effect—vote twice.

## 3.4 Comparison to bagging for classification

Microarray data, such as the ones arising in gene expression data, lend themselves to analysis with the objective of classifying future observations; in other words, using the data to decide if a future observation belongs to the control or the patient group—the decision being based on the new observation's 'features' (i.e. gene expressions) only. Since Breiman's (1996) original bagging was aimed at improving predictors and classifiers, it is of no surprise that there is already a body of literature on bagging and subagging microarrays with the purpose of classification; a partial list includes Dettling (2004), Dudoit and Fridlyand (2003), and Dudoit, Fridlyand and Speed (2002).

Although related at the outset, classification is a very different problem than hypothesis testing; their objectives are quite different, and so are the methods involved. To illustrate this point, we now give a brief description of the bagging/subagging procedures as used for microarray classification.

To start with, concatenate the $X$ and $Y$ matrices into a big $N \times n$ matrix denoted by $W$ where $n = n_X + n_Y$. Let $\underline{w}_1, \ldots, \underline{w}_n$ denote the columns of $W$, and define new variables $U_1, \ldots, U_n$ such that $U_i = 0$ for $i \leq n_X$, and $U_i = 1$ for $i > n_X$; in this sense, the variable $U_i$ is an indicator of which group (normal or patient) the $i$th subject belongs to. Finally, define $Z_i = (\underline{w}_i, U_i)$ for $i = 1, \ldots, n$.

The $Z_i$ data are multivariate but they constitute a single sample. This sample can be bootstrapped—by sampling with replacement from the set $\{Z_1, \ldots, Z_n\}$, or subsampled—by sampling without replacement from the same set $\{Z_1, \ldots, Z_n\}$, in order to create (pseudo)samples. In all the above-referenced works, bagging/subagging for microarray classification follows the above paradigm.

Note, however, that the above single-sample bootstrap scheme can generate (pseudo)samples that are unbalanced in terms of the two groups (normal/patient). To elaborate, let $Z_i^* = (\underline{w}_i^*, U_i^*)$ for $i = 1, \ldots, n$ be the bootstrap (pseudo)sample. Then, it is not unlikely that $\sum_{i=1}^n U_i^*$ turns out quite different from its expected value of $n_Y$; in fact, it is even possible (although very unlikely) that $\sum_{i=1}^n U_i^*$ is 0 or $n$, i.e., the (pseudo)sample consisting of data from one group only.

The above discussion refers to bootstrap and bagging but similar ideas hold for single-sample subagging. Let us define a (pseudo)sample to be balanced if the proportion of patients to control subjects within the (pseudo) sample is equal to that found in the original sample, i.e., $n_Y/n_X$. If we let $Z_i^* = (\underline{w}_i^*, U_i^*)$ for $i = 1, \ldots, b$ be the subsampling (pseudo)sample, then it is still possible to have $\sum_{i=1}^n U_i^* = 0$ provided of course that $b \leq n_X$. But even barring such extreme events, it is clear that there is no guarantee that the above subsampling (pseudo)sample would be balanced.

In conclusion, the possibility of unbalanced (pseudo)samples might not adversely influence the properties of bagging/subagging for classification purposes but it is problematic in our hypothesis testing setting. The balanced bagging/subagging procedures of Section 3.2 are devoid of this deficiency, since they yield—by design—exactly balanced (pseudo)samples.

Finally, note that different resampling methods have been used in connection with multiple comparisons—the most popular of which involving permutation tests; see e.g. Westfall and Young (1993), Ge, Dudoit and Speed (2003), or Romano and Wolf (2004). Nevertheless, the approach of Section 3.2 constitutes the first—to our knowledge—application of the notion of bagging/subagging to the problem of multiple comparisons.

An application and comparison (by simulation) of the proposed bagging/subagging procedures is given in:

www.math.ucsd.edu/~politis/PAPER/Bagging.pdf

where bagging and subagging are shown to indeed succeed in improving the experiment's discovery power at a small cost in increased FDR; as expected, 'maximum contrast' subagging appears to have an edge over bagging.

## REFERENCES

[]Benjamini, Y. and Hochberg, Y. (1995). Controlling the false discovery rate: a practical and powerful approach to multiple testing, J. Roy. Statist. Soc., Ser. B, 57, 289-300.

[]Breiman, L. (1996). Bagging predictors. Machine Learning, 24, 123-140.

[]Bühlmann, P. (2003). Bagging, subagging and bragging for improving some prediction algorithms, Recent Advances and Trends in Nonparametric Statistics, (M.G. Akritas and D.N. Politis, Eds.), Elsevier (North Holland), pp. 19-34.

[]Bühlmann, P. and Yu, B. (2002). Analyzing bagging, Ann. Statist., 30, 927-961.

[]Dettling, M. (2004). BagBoosting for tumor classification with gene expression data, Bioinformatics, vol. 20, no. 18, 3583-3593.

[]Dudoit, S. and Fridlyand, J. (2003). Bagging to improve the accuracy of a clustering procedure, Bioinformatics, vol. 19, no. 9, 1090-1099.

[]Dudoit, S. Fridlyand, J. and Speed, T. (2002). Comparison of discrimination methods for the classification of tumors using gene expression data, J. Amer. Statist. Assoc., vol. 97, no. 457, 77-87.

[]Efron, B. (1979). Bootstrap methods: another look at the jackknife, Ann. Statist., 7, 1-26.

[]Efron, B. (2005). Local false discovery rates, Preprint available from: http://www-stat.stanford.edu/~brad/papers.

[]Efron, B. (2006). Size, power, and false discovery rates, Preprint available from: http://www-stat.stanford.edu/~brad/papers.

[]Efron, B. and Tibshirani, R.J. (1993), An Introduction to the Bootstrap, Chapman and Hall, New York.

[]Ge, Y., Dudoit, S. and Speed, T. (2003). Resampling-based multiple testing for microarray data analysis, Test, vol. 12, no. 1, 1-77.

[]Miller, R.G. (1981). Simultaneous Statistical Inference, 2nd Ed., Springer, New York.

[]Newton, M.A., Noueiry, A., Sarkar, D., and Ahlquist, P. (2004). Detecting differential gene expression with a semiparametric hierarchical mixture method Biostat., 5, 155-176.

[]Politis, D.N., Romano, J.P. and Wolf, M. (1999), Subsampling, Springer Verlag, New York, 1999.

[]Romano, J.P. and Wolf, M. (2004). Exact and approximate stepdown methods for multiple hypothesis testing. Journal of the American Statistical Association, 100, 94-108.

[]Shao, J. and Wu, C.F. (1989). A general theory of jackknife variance estimation, Ann. Statist., 17, 1176-1197.

[]Singh, D., Febbo, P.G., Ross, K., Jackson, D.G., Manola, J., Ladd, C., Tamayo, P., Renshaw, A.A., D'Amico, A.V., Richie, J.P., Lander, E.S., Loda, M., Kantoff, P.W., Golub, T.R., and Sellers, W.R. (2002). Gene expression correlates of clinical prostate cancer behavior, Cancer Cell, vol. 1, no. 2, 203-209.

[]Tukey, J.W. (1958). Bias and confidence in not quite large samples, (Abstract) Ann. Math. Statist., 29, 614.

[]Tusher, V.G., Tibshirani, R., and Chu, G. (2001). Significance analysis of microarrays applied to the ionizing radiation response, Proc. Natl. Acad. Sci. U.S.A., 98, 5116-5121.

[]Westfall, P. and Young, S. (1993). Resampling-based multiple testing: examples and methods for $p$-value adjustment, Wiley, New York.